\documentclass[twocolumn, aps,preprintnumbers,amsmath,amssymb,nofootinbib,prl]{revtex4-1}

\usepackage{color}
\usepackage{graphicx}
\usepackage{epsfig}
\usepackage{textcomp}
\usepackage{slashed}

\newcommand{\me}{\slashed E}

\def\lsim{\mathrel{\rlap{\lower4pt\hbox{\hskip1pt$\sim$}}
    \raise1pt\hbox{$<$}}}         
\def\gsim{\mathrel{\rlap{\lower4pt\hbox{\hskip1pt$\sim$}}
    \raise1pt\hbox{$>$}}}         
                           
\begin{document}

\preprint{\today}

\title{$W'$ in new physics models at the LHC}

\author{DongHee Kim and Youngdo Oh}
\affiliation{WCU High Energy Collider Physics Research, Kyungpook National University, \\Sankyuk-dong, Buk-gu, Daegu, 702-701, Korea}
\author{Seong Chan Park}
\email{sc@jnu.ac.kr}
\vspace{1cm}
\affiliation{Department of Physics, Chonnam National University, 300 Yongbong-dong, Buk-gu, Gwangju, 500-757, Korea
}
\vspace{1.0cm}

\begin{abstract}
We study the new heavy charged gauge boson $W'$ in various models including Left-Right symmetric, Little Higgs,  Randall-Sundrum and universal extra dimension model considering $pp \to W' \to \ell \nu_\ell$ with $\sqrt{s}=7$ TeV and $14$ TeV at the LHC. Of particular, we show that the universal extra dimension model is highly constrained by existing and forthcoming data.

\end{abstract}


\maketitle

\newpage

\section{Introduction} 
The LHC has started to probe the unprecedented domain of physics above a TeV scale. One of the important goals in this domain is to find a new force beyond the standard strong and electroweak force of the gauge symmetry. The new force may be originated from the extended gauge symmetry or deeper structure of spacetime. 
Depending on the specific extension of the Standard Model (SM), there could exist new gauge bosons, namely $Z'$, a neutral gauge boson  and $W'$, charged ones. Indeed several new physics models, including Left-Right(LR) symmetric, Randall-Sundrum, Little Higgs and universal extra dimension(UED) model, predict $Z'$ and $W'$ or both. 
$W'$, in particular, leaves a rather clean signature of lepton plus missing energy ($\me$), which is supposed to be from $\ell \bar{\nu_\ell}$ or  $\bar{\ell} \nu_\ell$ \footnote{The $Z'$ search using $\ell \bar{\ell}$, which we leave for the future study,  is also important and can be complementary to the current $W'$ search depending on models. In large extra dimension models, Spin-2 heavy graviton, $G$,  can also produce $\ell \bar{\ell}$ so that the discrimination between $Z'$ and $G$ is another challenging issue for the LHC.}. 
So far these channels have been searched to look for $W'$ expected in LR model with the hadron collisions\cite{CDFpaper, CMSpaper}.  However the search can be applied to other models. One can use independent limits on $\sigma(pp\to W') Br(W'\to \ell \nu_\ell)$   by  LHC experiments to set bounds or to discover the particle from the models. Of particular, the UED models could be one of the well motivated TeV scale new physics models offering interesting signatures at the LHC.

UED models are based on an extended spacetime 
in five dimensions with the coordinate $x^M = (x^\mu, x^4 )$ where $x^\mu$ is for  large dimensions with $\mu=0,1,2,3$ and the fifth dimension is compact as $x^4 \in \left[-L, L\right] = [-\pi R, \pi R]$ \cite{UEDoriginal}. All the matter fermions (quarks and leptons) and gauge bosons ($g, W,  B$) corresponding to  the SM gauge group $SU(3)_c\times SU(2)_W \times U(1)_Y$ are propagating through the  5D bulk of the extended dimension so that their Kaluza-Klein(KK) states, $q_n, \ell_n, \nu_n, g_n, W_n, Z_n$ and  $\gamma_n$, where $n$ denotes the $n$-th KK excitation, may give rise to  various new phenomena at the LHC provided that the size of extra dimension is within the reach of the LHC: $1/R \lsim 10$ TeV \cite{Antoniadis}. By the exact $\mathbb{Z}_2$ reflection symmetry about the middle point of the extra dimension $x^4=0$, dubbed as KK-parity, the essential features of the UED  phenomenology at the collider experiment are closely parallel to the ones in supersymmetric models with the exact R-parity conservation \cite{bosonic SUSY}.  Consequently  UED models provide a nice dark matter candidate: the lightest KK-parity-odd particle (LKP)  \cite{KKDM1, KKDM2} preferably KK-photon \cite{Cheng} \footnote{Precise relic density calculations of the KK dark matter has been done \cite{KKDM3, KKDM4}. For a recent summary of the minimal realization of UED model and computation, see, e.g.\cite{mUED}. Also see \cite{Matsumoto:2007dp} for a scenario with right handed neutrinos.}.

The level-2 KK gauge bosons, $\gamma_2, Z_2$ and $W_2$ are particularly interesting  as they can directly couple with the SM fermions. On the other hand, the level-1 KK bosons, even though they are lighter than the level 2-KK bosons, are not allowed to couple with the SM fermions due to the KK-parity conservation, which is essential for providing dark matter candidates. Indeed, the processes induced by odd-KK states $ qq' \to W_{2n+1} \to \ell \nu_\ell$ or $ q\bar{q} \to \gamma_{2n+1}, Z_{2n+1} \to \ell \bar{\ell}$ are all forbidden by KK-parity but  the processes through the even KK gauge bosons $qq' \to W_{2n} \to \ell \nu_\ell$ or $q\bar{q}\to \gamma_{2n}, Z_{2n} \to \ell \bar{\ell}$ are allowed at one-loop level \cite{bosonic SUSY,Cheng} or at the tree level \cite{sued1,sued2,sued3,sued4,Csaki}.  The other decaying channels to low state gauge bosons: $W_2 \to Z_0 W_0, Z_1W_1$, etc are all suppressed by a factor of weak mixing angle for KK-states $(m_W R)^2 \ll 1$ and/or tiny phase space on top of the loop-suppression factor ($g^2/4\pi$)  so that we will concentrate on the channels to the fermions  in the present study.
Particular attention is given to the direct coupling of the level-2 KK W-boson to the SM fermions (i.e. $u \bar{d}$ and $\ell \nu_\ell$) which is allowed when the bulk fermion mass for the fermions is non-vanishing \cite{sued1,sued2,sued3,sued4,Csaki}.  

The paper is organized as follows:  we briefly review $W'$ in LR, LH, RS, UED models and  the most relevant properties of KK W-bosons in UED models for the LHC search.  Then we calculate the production rate as well as decay width of KK W-boson and study the perspectives at the LHC.

\section{$W'$ in various physics models}
In this section we examine the candidates of $W'$ bosons in various new physics models which can be potentially tested at the LHC by $\ell +\me$ signature.

In a Left-Right symmetric model with an extended gauge symmetry, $SU(2)_L\times SU(2)_R$, a new gauge boson, $W_R$, can be a candidate of $W'$ \cite{LRmodel} . Here the right-handed fermion doublets $Q_R=(u,d)_R$ and $L_R=(\nu, l)_R$ couples to $W_R$ just like the left-handed corresponding states in the SM to the conventional $W$-boson. The $SU(2)_R$ symmetry is broken (by a new Higgs boson of the $SU(2)_R$ symmetry) at some high scale $\Lambda \sim m_{W_R}$. Since $W_R \to u_R d_R$ or $W_R\to e_R \nu_R$ are allowed, the LHC may observe $e_R +\me$ signatures if $m_{\nu R}$ is light enough. If $\nu_R$ has a large Majorana mass thus is significantly heavier than $\nu_L$ the signature of $e_R + \me$ can be also modified. This possibility has been examined recently by Tevatron and the LHC with electrons.

In Randall-Sundrum model \cite{RS}, gauge bosons may propagate through the five dimensional bulk like in UED models so that (odd) KK-states of gauge bosons can give rise to $\ell +\me$ signatures. However, the coupling strength between $W_{KK}$ and light fermions is model dependent. Actually the light fermions tend to localize toward the UV-brane producing the reduced values of Yukawa couplings with the Higgs boson on IR-brane but  the KK-gauge bosons lean toward the IR brane so that the resultant waver function overlap or the effective coupling between the KK-boson and light fermions is quite small. \footnote{The localization is controlled by a bulk Dirac mass.\cite{Grossman}}  On the other hand, a TeV-scale graviton can be produced and decay to lepton pairs or jets  in this model which may provide a handle to discriminate different models. 

In little Higgs (LH) models \cite{LH1,LH2,LH3},  the Higgs field is a pseudo-Goldstone boson and its potential energy is generated by so-called `collective symmetry breaking' mechanism which requires an extension of the standard model gauge symmetry. Resultantly, there can be a new charged gauge boson ($W'\sim W_R$). However, there are issues regarding the electroweak precision observables \cite{LH_Big} so that a realistic model needs some sophisticated modifications. Avoiding some of such a problem, a $Z_2$ parity, dubbed T-parity is introduced, under which the new gauge boson has an odd parity but  the SM fermions have even parities so that the direct decay of the new boson to the SM fermion pair ($W_R\to \ell \nu_\ell$) is forbidden \cite{LHT1,LHT2}.

In UED models, particles propagate in higher dimensional bulk and so have their KK excited modes. The zero modes correspond to the standard model particles. The second  KK state of $W$-boson or $W_2$ can decay to $\ell + \nu_\ell$, which are the zero modes, even though the first KK state, $W_1$, which is lighter than $W_2$, cannot directly decay to the standard model particles due to the KK-parity conservation. The strength of the $W_2-\ell-\nu_\ell$ interaction is determined by the wave function overlap and can be sizable. 

In summary, among well-motivated various new physics models which predict a new charged gauge boson $W'$,  UED with bulk mass of fermion has the most probable chance to get observed by the direct measurement of $\ell+\me$ at the LHC. We thus focus on UED model in detail below.


\section{The $W_{\rm KK}$ bosons in UED models}


In UED models, the gauge bosons of the gauge symmetry $SU(3)_c \times SU(2)_W \times U(1)_Y$ contains the zero modes which correspond to the gluon, $g$, weak gauge bosons, $W, Z$, and photon, $\gamma$, of the SM and their KK excited states, $g_n, W_n, Z_n$ and $\gamma_n$, here $n$ is positive integer number. We are mostly interested in $W_n$ in this paper. The $n$-th excited $W$-boson has the mass:
\begin{eqnarray}
m_{W_n}^2 \equiv m_n^2 = m_W^2 + \left(\frac{n}{R}\right)^2.
\label{eq:m2}
\end{eqnarray}

With the exact KK parity in UED models, the direct couplings of the KK-odd gauge bosons $W_{n=1,3,5,..}$ with the pair of the SM fermions, i.e., the zero modes of the bulks fields, $f_0 =\left(e, \mu, \tau, u, d, ..\right)$  are all forbidden: $g_{W_{2n+1}-f_0-f'_0}=0$. On the other hand, KK-parity does allow the KK-even gauge boson-SM fermion couplings, $W_{2n}-f_0-f'_0$,  in general. Effectively, the gauge couplings of $W_n$ with the SM fermions can be written
\begin{eqnarray}
g_n = g^{\rm SM} {\cal{F}}_n (\mu_\psi L),
\end{eqnarray}
where ${\cal F}_n$ is the wave function overlaps between the $n$-th KK gauge boson and the SM fermions: ${\cal F}_n\equiv \int dx^4 f_0 f_0 f_{W_n}$ as is defined  in Ref. \cite{sued1} , which depends on the bulk mass parameter of the corresponding fermion $\mu_\psi $. 
Thanks to the KK parity conservation, ${\cal F}_{odd}=0$ \footnote{At the limit of  vanishing bulk mass, $x =\mu R \pi \to 0$, mUED recovered where not only KK-parity but also KK-number is conserved at tree-level. However, a small higher order correction can induce a loop-suppressed KK-number violation. For $W_2\to \ell \nu_\ell$, for instance, the value is found \cite{mUED}:
${\cal{F}}_2(x\to 0) \approx \frac{1}{\sqrt{2}} (\frac{9}{8} g_1^2 - \frac{33}{8} g_2^2)
 \frac{1}{16\pi^2}\log (\Lambda R)^2 \approx -0.04$ with $\Lambda R =20$, which we will neglect in the paper.} and ${\cal F}_{even}\neq 0$, in general:
\begin{equation}
{\cal F}_n(x)=\begin{cases}
 0& \text{ if } n=2m+1 \\ 
 \frac{x^2 (-1+(-1)^m e^{2x})(\coth x-1)}{\sqrt{2(1+\delta_{m0})}(x^2 + m^2\pi^2/4)}& \text{ if } n= 2m . 
\end{cases}
\label{eq:fn}
\end{equation}
%
Note that ${\cal F}_0 =1$ for an arbitrary $x$ meaning $g_0 = g^{\rm SM}$ as is obviously expected. Another interesting limit is $x\to \infty$ where ${\cal F}_{2n}\to (-1)^n \sqrt{2}$.


\section{$ q\bar{q'} \to W_{2n} \to \ell \nu_\ell$ in UED}
\label{sec:xsection}

The parton level cross section for $u \bar{d} \to W \to \ell \nu_\ell$ scattering through the zeroth state $W$ at the tree level approximation is
\begin{eqnarray}
\frac{d \hat{\sigma}_{\rm SM}}{ d \Omega}=\frac{g^4 |V_{ud}|^2}{48 \cdot (4\pi)^2}\frac{\hat{u}^2}{\hat{s}} |\Pi_{\rm SM}(\hat{s})|^2,\\
\Pi_{\rm SM}(\hat{s})= \frac{1}{\hat{s}-m_W^2-i m_W \Gamma_W(\hat{s})}.
\end{eqnarray}

On the other hand, including a tower of KK-$W_n$ gauge bosons, we get the cross section for $u\bar{d}\to W_{\rm KK} \to e \nu$ : 
\begin{eqnarray}
\frac{d \hat{\sigma}_{\rm UED}}{ d \Omega}=\frac{g^4 |V_{ud}|^2}{48 \cdot (4\pi)^2}\frac{\hat{u}^2}{\hat{s}} |\Pi_{\rm UED}(\hat{s})|^2, \\
\Pi_{\rm UED}(\hat{s},\mu, 1/R)=\sum_{n} \frac{{\cal F}_n^2}{\hat{s}-m_n^2-i m_n \Gamma_n(\hat{s})},
\end{eqnarray}
where only even states with $n=0, 2, 4, \cdots$ can contribute. $m_n$ and $\Gamma_n$ are the mass and the decay width of $W_n$, respectively.  It is noticed that the cross section for a given parton level CM energy $\hat{s}$ is obtained by scaling the SM result corresponding to the case with $n=0$:
\begin{eqnarray}
\frac{\hat{\sigma}_{\rm UED}}{ \hat{\sigma}_{\rm SM}}=\frac{|\Pi_{\rm UED}|^2}{|\Pi_{\rm SM}|^2}(\mu,1/R)
\end{eqnarray}
where we assumed a universal bulk mass parameter $\mu$ for all the bulk fermions.
If the ratio is significantly different from unity, the LHC may have a chance to detect the signature of UED models on which we will consider more carefully in the next section.  

\subsection{$\Gamma(W_{2n} \to \ell \nu_\ell)$}
For $W\to e \nu$, the partial decay width of $W$ is given with a small higher order correction $\delta\ll 1$:
\begin{eqnarray}
\Gamma(W_{n}\to e \nu) = \frac{g_{n}^2}{48\pi} m_{n} (1+\delta),\,\,n=0,2,4,\cdots
\end{eqnarray}
$\Gamma(W_0\to e\nu) \approx 226.5\pm 0.3$ MeV with $\delta_{SM}\lsim 0.5\%$. There are 3 leptonic decay channels to $\ell \nu_\ell$, $\ell=e,\mu,\tau$. Also there are hadronic decay channels to $q \bar{q'}$ where $q=u,c (t)$ and $q'=d',s'(b')$ with the possible decay channel to the third generation quarks for heavier excited modes ($n\geq 1$).  Here we neglect the chance of decaying to level-1 KK particles, e.g., $W_{2}\to \ell_1 \nu_1, W_1 Z_1$ etc, taking the heavier KK-fermion masses and the tiny available phase spaces.

Including  QCD corrections, we found the total decay width can be well approximated by a simple formula
\begin{eqnarray}
\Gamma_{W_n} \approx \Gamma(W_n\to e \nu)\left(N_f + \tilde{N}_f  N_c (1+{\cal O}\left(\frac{\alpha_s}{\pi}\right) \right)
\end{eqnarray}
where $N_f=3$ and $N_c=3$ without introducing an exotic flavor states. $\tilde{N}_f =3-\delta_{n0}$ considering the third generation quarks. 
\begin{eqnarray}
\Gamma_0 \approx 9 \Gamma(W\to e\nu),\\
\Gamma_{n\geq 1} \approx 12 \Gamma(W_n \to e\nu). 
\end{eqnarray}

A useful formulae is found:
\begin{eqnarray}
\frac{\Gamma (W_n \to e\nu)}{\Gamma (W\to e \nu)} \approx \frac{m_n}{m_W}{\cal F}_n^2  \approx \frac{n {\cal F}_n^2}{m_W R}
\end{eqnarray}
assuming that $\delta_n \sim \delta_0 \ll 1$ and $1/R \gg m_W$.

\section{The LHC bounds on the second KK W boson ($W_2$) }
\label{sec:LHC}

\begin{figure}[ht]
\centering
\includegraphics[width=0.48 \textwidth]{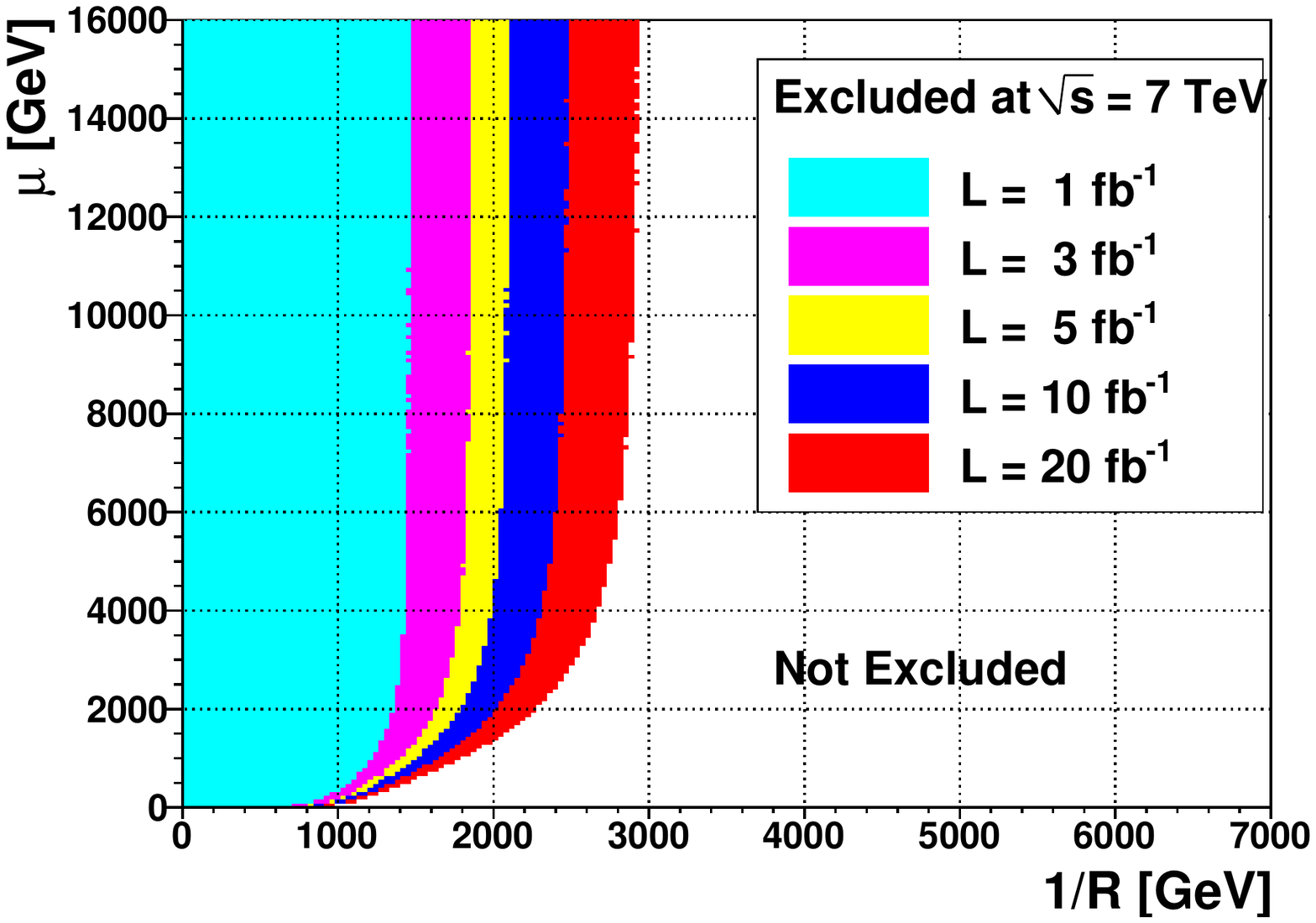}
\caption{7 TeV sensitivity limits on UED parameter space in $(1/R,\mu)$ plane. Expected sensitivity regions by the LHC with the luminosity corresponding to $1,3,5,10$ and $20 ~{\rm fb^{-1}}$ (from left to right colored regions), respectively, are plotted. }
\label{fig:7TeV}
\end{figure}

\begin{figure}[h]
\centering
\includegraphics[width=0.49\textwidth]{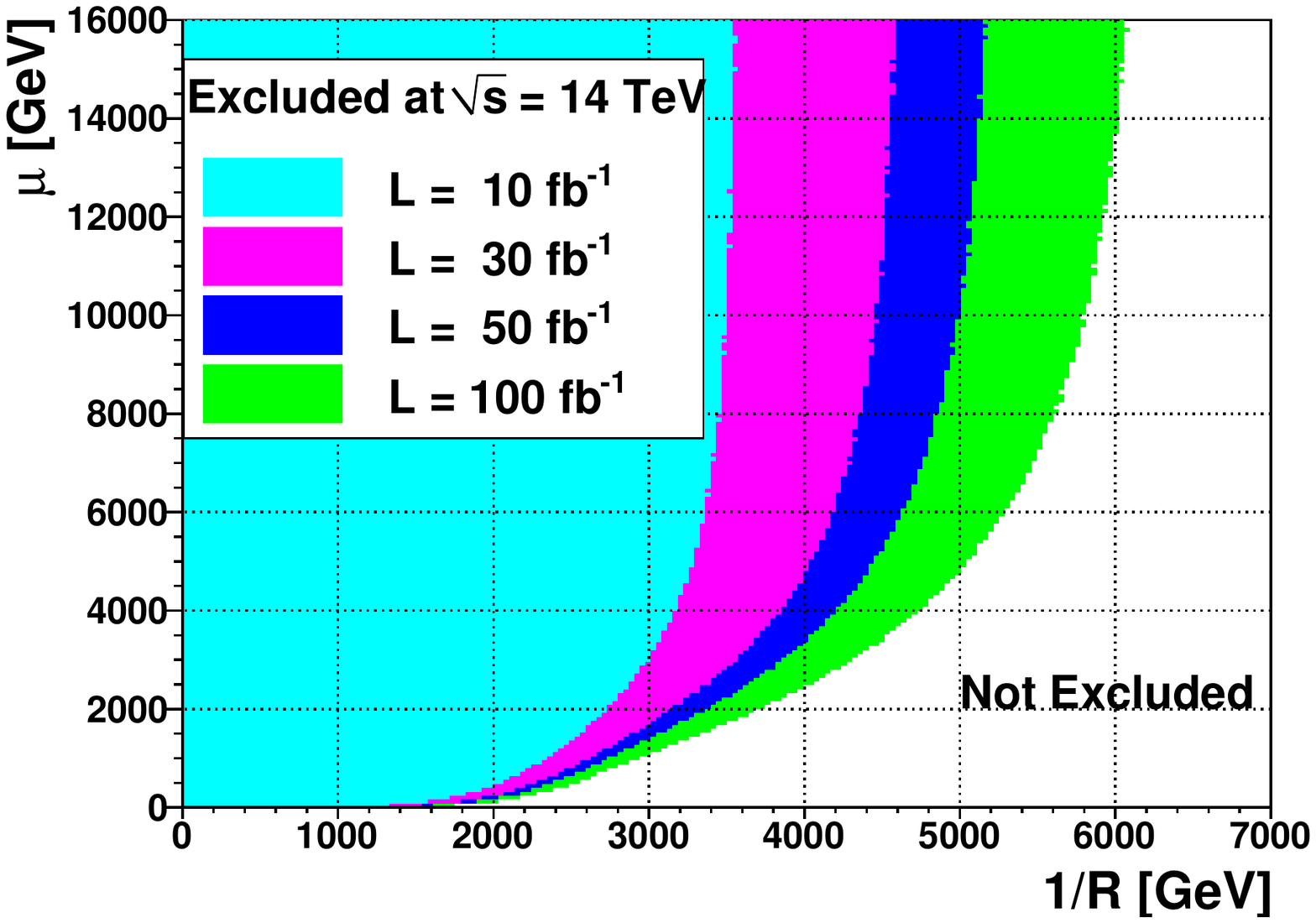}
\caption{Same plot as in Fig. \ref{fig:7TeV} with a higher energy,14 TeV, run for the  expected limit on UED parameter space in $(1/R,\mu)$ plane. Expected sensitivity regions by the LHC with the luminosity corresponding to $10, 30, 50$ and  $100 ~{\rm fb^{-1}}$ (from left to right colored regions), respectively, are plotted. }
\label{fig:14TeV}
\end{figure}

The $W_2$ production cross section from $pp$ collision is calculated using the 
{\sc pythia} Monte carlo event generator. 
The masses and new coupling constants $W_{2n}$ are calculated using formula 
\eqref{eq:m2} and \eqref{eq:fn}, respectively. Then those numbers are passed to the {\sc pythia}
as input parameters.  
The CTEQ6L1 parton distribution function is used for PDF convolution.
Finally, the cross section times branching ratio of electron channel at LO level
is calculated at center of mass energy  7 TeV and 14 TeV.
The scaned UED paramter space is $1/R = [0, 7000 {\rm GeV}]$
and $\mu = [0, 16000 {\rm GeV}]$. The calculated cross sections are compared with
the 95\% confidence level(C.L.) cross section limit for the various luminosities
to make sensitivity regions on the UED parameter space.
The 95\% C.L. cross section limit is
obtained from 3/luminosity with the assumption which the observed
signal event is 0 and detector efficiency is 100\%.  Figure \ref{fig:7TeV} and \ref{fig:14TeV} show
the sensitivity regions  at $\sqrt{s}$ = 7 TeV and 14 TeV respectively.

\section{Conclusion}
\label{sec:conclusion}

We study the new heavy charged gauge boson $W'$ in various models considering $pp \to W' \to \ell \nu_\ell$ with $\sqrt{s}=7$ TeV and $14$ TeV at the LHC.  Of particular, we show that the universal extra dimension ``KK dark matter" model is highly constrained by existing and forthcoming data. We consider the charged second KK gauge boson production at the LHC, which provides Lepton+$\me$ signatures in universal extra dimension. 
Assuming forthcoming data corresponding to $5(20)\text{fb}^{-1}$ of $\sqrt{s}=7 $~ TeV, we show that the LHC can cover a rather large range of Kaluza-Klein scale up to $1/R \lesssim 2 (3)$ TeV, respectively for various value of $\mu$.  Even higher scale $1/R \lesssim 3.4 (6.0)$ TeV is within the reach of  higher energy run with $\sqrt{s}=14$  TeV and integrated luminosity ${\cal L}=10 (100) \text{fb}^{-1}$, respectively. Since a naturally required scale for KK-photon dark matter is $1/R\lesssim \text{a few}$ TeV \cite{KKDM3, KKDM4}, we conclude that LHC can essentially exclude or prove UED model as a model of dark matter.

\vspace{0.5cm}
{\bf Acknowledgement:} This work was supported by the World Class 
University(WCU) Program(R32-2008-000-20001-0) and Basic Science Research Program(2011-0010294) through the National Research Foundation(NRF) of Korea funded by the Ministry of Education, Science and Technology.

\end{document}